\documentclass[twocolumn]{revtex4}
\usepackage[utf8]{inputenc}
\usepackage{graphicx}
\usepackage{color}

\begin{document}

\title{Dielectric response of the antiferromagnetic multiferroics with cycloidal equilibrium for the polarization related to the noncollinear spins}

\author{Pavel A. Andreev}
\email{andreevpa@my.msu.ru}
\affiliation{Department of General Physics, Faculty of physics, Lomonosov Moscow State University, Moscow, Russian Federation, 119991.}

\date{\today}

\begin{abstract}
Linear dynamics of the small amplitude perturbations in the antiferromagnetic multiferroics is analytically considered.
Equilibrium state in the considered regime is assumed to be cycloidal structure both for the antiferromagnetic vector and the nonzero magnetization vector
(presence of weak ferromagnetism in the system is included).
The electric polarization of spin origin related to the "noncollinear" spins is considered and its influence on the spin dynamics is included.
The dynamical electric susceptibility tensor is obtained to demonstrate response of the multiferroics on the alternating electric field.
Existence of two polarization projections is found, response of each of them on two projections of the electric field perpendicular to the cycloid direction is presented.
Different roles of the in plane and out of plane electric perturbations is demonstrated analytically.
Essential role of the low frequency gapless spin wave in the dielectric response is described.
Contributions of the spiral structure and the weak ferromagnetism in the dielectric response are specified.
\end{abstract}


\maketitle



\section{Introduction}

Linear response on the small amplitude perturbations of various materials still essential
nowadays,
especially for the multiferroic materials demonstrating the magnetoelectric effect.
One of recent examples of such phenomena is the electromagnons
\cite{Pimenov NP 06}, \cite{ShuvaevPimenov EPJB 11}, \cite{Pimenov JP CM 08}, \cite{ElsasserPimenov NJP 17},
which is also described in reviews \cite{Tokura RPP 14}, \cite{Dong AinP 15}.
A possible mechanism for the electromagnon explanation is suggested in Ref.
\cite{Katsura PRL 07},
while some recent theoretical papers address this phenomenon
in its basic features \cite{Castro PRB 25}, \cite{AndreevTrukh EPL 25}.
Both papers \cite{Castro PRB 25}, \cite{AndreevTrukh EPL 25} discuss the magnetoelectric effect
and corresponding collective excitations for the ferromagnetic materials,
while the experimental evidence are found for the antiferromagnetic multiferroics
\cite{Dong AinP 15}.
Regarding electromagnons we are interested in the magnetoelectric effect of spin origin \cite{Tokura RPP 14}.
It leads to necessity to analyze the dielectric response of the antiferromagnetic materials (AFM)
related to the purely spin dynamics.
Especially, it is important to distinguish the influence of the cycloidal spiral on the dielectric response.
It is also essential to specify the role of noncollinearity of two spin sublattices in the cycloidal AFM
\cite{Dong AinP 15},
showing the weak ferromagnetism of the AFM samples usually related to the Dzylaoshinskii-Moriya interaction.

There are several manifestations of the magnetoelectric effect of spin origin
\cite{Tokura RPP 14}, \cite{Fiebig JP D 05}, \cite{Mostovoy npj 24}.
Their classification changes over the time towards the decrease of number of mechanisms,
so three mechanism can be mentioned following review \cite{Tokura RPP 14}.
Additional mechanisms is recently suggested in Ref. \cite{Andreev 2025 12} for the AFM multiferroics.

The role of the spiral equilibrium structure on the spin dynamics is addressed in literature
\cite{Fishman PRB 19}, \cite{Andreev 2025 09}, \cite{Andreev 2025 10}.
Extensive analysis of the spin-wave dispersion in multiferroic TbMnO$_{3}$ is given in Ref. \cite{Holbein PRB 23},
where the detailed description of the magnetic equilibrium structures and the multiferroic equilibrium order in TbMnO$_{3}$ is given as well.
The illustration of the crystallographic unit cell of TbMnO$_{3}$,
showing Mn moments in the spin spiral phase,
can be found in Ref. \cite{Holbein PRB 23} (see Fig. 3 therein).
One direction shows up-up-down-down order as a part of more complex structure.
The spin wave dispersion dependencies for the model quasi-one-dimensional structure can be found in Ref. \cite{Andreev 2026 03}.
Particularly, authors assume that the polarization has the spin nature
and it is is associated with the noncollinear parts of spins (see eq. 1).
Three low-energy modes are measured at the magnetic zone center in Refs. \cite{Senff PRL 07}, \cite{Senff JP CM 08}
for the in multiferroic TbMnO$_{3}$ being in the cycloidal phase.
These modes are an in-plane mode (with respect to the cycloidal bc plane)
and two out-of-plane modes.

Low-energy optical spectrum in the multiferroic RMnO$_{3}$,
where for the origin of the electromagnon excitation, the coupling between the external electric
field and the spin-dependent electric polarizations (associated with the collinear parts of spins)
is considered in Ref. \cite{Mochizuki 1001}.
Here the form of magnetoelectric coupling differs from mentioned above,
but above we mentioned mechanism for the equilibrium,
while here authors deal with the dynamical response.

Spin-wave and electromagnon dispersions in multiferroic MnWO4 are described experimentally and theoretically in
\cite{Xiao PRB 16},
where 12 different nearest neighbors are included for the three dimensional structures.
Spin waves are also considered in skyrmion crystals \cite{Hirosawa PRX Q 22}.

\section{Model}

\subsection{Geometry of the samples}

We consider the AFM with the cycloidal spiral equilibrium structure
both for the antiferromagnetic vector $\textbf{L}=\textbf{S}_{A}-\textbf{S}_{B}$,
and the "magnetization" vector $\textbf{M}=\textbf{S}_{A}+\textbf{S}_{B}$
We work with the spin densities,
so it might be incorrect to use the notion "magnetization" vector,
while we dial with the full spin density, which is proportional to the magnetization of the sample.
Vectors $\textbf{S}_{A}$ and $\textbf{S}_{B}$ are the partial spin densities of two magnetic sublattices composing the AFM.
We also note $\mid\textbf{S}_{A}\mid=\mid\textbf{S}_{B}\mid$.

Explicit form of the equilibrium vectors is
\begin{equation}\label{MEEinAFMnc L eq cycloid}
\textbf{L}_{0}=L_{b}\cos(qx)\textbf{e}_{x}+L_{c}\sin(qx)\textbf{e}_{y}
, \end{equation}
and
\begin{equation}\label{MEEinAFMnc M eq cycloid}
\textbf{M}_{0}=M_{b}\sin(qx)\textbf{e}_{x}-M_{c}\cos(qx)\textbf{e}_{y}
, \end{equation}
where we see $(\textbf{L}_{0}\cdot\textbf{M}_{0})=0$,
it corresponds to $\mid\textbf{S}_{A,0}\mid=\mid\textbf{S}_{B,0}\mid$.
As an example of such structure mention Ref. \cite{Dong AinP 15} (see Fig. 3 as an illustration).
It also contains the spatial period of the cycloid $\lambda=2\pi/q=64$ nm.
However, comparison of this equilibrium structure with the form of crystallographic unit cell of TbMnO$_{3}$,
showing Mn moments in the spin spiral phase \cite{Holbein PRB 23},
shows the simple model nature of the structure (\ref{MEEinAFMnc L eq cycloid}), (\ref{MEEinAFMnc M eq cycloid}) considered in this paper.
Nevertheless, it is essential to give the analytical trace of different contributions in the dynamical response for such model system.

Formation of nonzero equilibriums magnetization $\textbf{M}_{0}$ is usually related
to the Dzylaoshinskii-Moriya interaction in the form related to two partial ligand shifts
(see for instance recent discussion of different forms of the Dzylaoshinskii-Moriya interaction in Ref. \cite{Andreev 2025 11},
and a shorted discussion can be found in Ref. \cite{Andreev 2026 07}),
which exists in AFM or other multicomponent magnetics.

The formation of the cycloid itself (for AFM vector $\textbf{L}_{0}$)
can be related to the Dzylaoshinskii-Moriya interaction as well,
but to its different form related to single partial ligand shift
(one of two partial ligand shifts mentioned above,
so both of them can exist simultaneously).
It is also described in Refs. \cite{Andreev 2025 11} and \cite{Andreev 2026 07}.

Effective spin-spin interaction related to the odd anisotropy of the symmetric
exchange interaction (OASEI), suggested in Ref. \cite{Andreev 2025 12},
can also contribute in the formation of noncollinear structure.

Here we do not consider the formation of the described
(\ref{MEEinAFMnc L eq cycloid}) and (\ref{MEEinAFMnc M eq cycloid})
equilibrium state,
which is well-known in literature
for description of the magnetization of some phases on known materials.
We are focused on the dielectric response of AFM multiferroics for the described equilibrium.
Moreover, we do not include the Dzylaoshinskii-Moriya interaction and OASEI in the evolution of perturbations
focusing on the exchange interaction and the anisotropy contribution.
The contribution of the magnetoelectric effect in the spin density evolution equation is included.
Being relatively small, it gives the coupling of the spin density to the electric field.

\begin{figure}\includegraphics[width=8cm,angle=0]{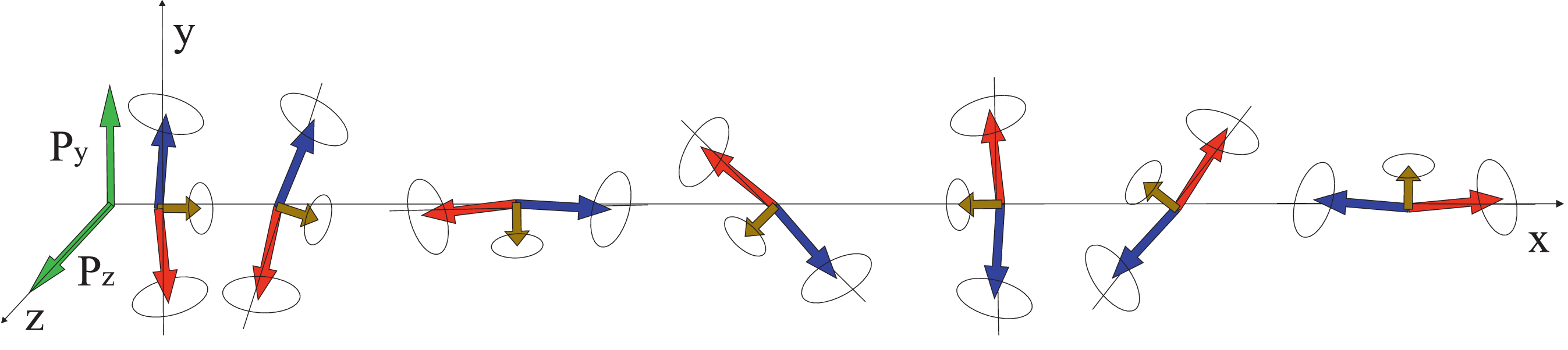}
\caption{\label{Fig 02}
(Color online)
The figure schematically shows the equilibrium AFM cycloid of "spin-up" (blue arrows) and "spin-down" (red arrows)
with the resulting "magnetization" vector (brown arrows)
existing due to the weak ferromagnetism caused by the Dzylaoshinskii-Moriya interaction.
The circles around arrows shows that we consider the small amplitude spin waves in this system.
On the left-hand side of the figure we plotted two green arrows signed as $\delta P_{y}$ and $\delta P_{z}$
and demonstrating the possible projections of the dynamical response of the electric polarization.}
\end{figure}

\subsection{Electric polarization}

In this paper we consider the electric polarization of spin origin related to the noncollinear parts of spins.
Examples of such model can be found in
Ref. \cite{Mostovoy PRL 06} (see eq. 2, for the macroscopic expression),
Ref. \cite{Tokura RPP 14} (see eq. 19, p. 8, for the atomistic expression),
Ref. \cite{Dong AinP 15} (see eq. 3, p. 533, for the macroscopic expression),
Ref. \cite{Mostovoy npj 24} (see eq. 10, p. 3, for the atomistic expression).

The electric dipole moment can be caused by the noncollinear parts of the neighboring spins.
The electric dipole moment in this regime is proportional to the vector product of these spins.
The vector product changes its sing at the renumeration of the spins.
The electric dipole moment should not depend on the numeration.
Therefore, the electric dipole moment contains the vector of relative position $\textbf{r}_{ij}$ of spins.
Finally, the electric dipole moment appears as
\begin{equation}\label{MEEinAFMnc edm operator NONsimm}
\hat{\textbf{d}}_{ij}= \alpha_{ij}
[\textbf{r}_{ij}\times[\hat{\textbf{s}}_{i}\times\hat{\textbf{s}}_{j}]],
\end{equation}
where $\alpha_{ij}$ is the coefficient depending on the module of relative position $\mid\textbf{r}_{ij}\mid=r_{ij}$ of spins,
(see for instance Ref. \cite{Mostovoy npj 24} eq. 10).
We consider the AFM hence we assume that the neighboring spins have opposite equilibrium spin direction
(predominantly opposite since we include the weak ferromagnetism in the sample).

The macroscopic electric polarization can be presented as
$$\textbf{P}=\frac{1}{6}g_{(\alpha),AB}\biggl\{\textbf{L}(\nabla\cdot\textbf{L})-(\textbf{L}\cdot\nabla)\textbf{L}$$
\begin{equation}\label{MEEinAFMnc P NONsimm}
-\textbf{M}(\nabla\cdot\textbf{M})+(\textbf{M}\cdot\nabla)\textbf{M}\biggr\}
, \end{equation}
where the interaction constant appears
$g_{(\alpha),AB}=\int r^2\alpha(r)d^3r$,
at the transition to the macroscopic scale.
This is generalization of the polarization,
well-known for the ferromagnetic materials
\cite{Tokura RPP 14}, \cite{Mostovoy npj 24}, \cite{Mostovoy PRL 06}, \cite{Baryakhtar JETP 83},
it can be also considered as a variation of AFM polarization considered in Ref. \cite{Sparavigna PRB 94}.

Derivation of polarization
can be addressed vis the spin-current model in a form based on the quantum hydrodynamic method
\cite{Andreev 2025 11},
\cite{AndreevTrukh PS 24},
\cite{AndreevTrukh AoP 26}.

Perturbations of the electric polarization in the linear on the small amplitude approximation has the following form
$$\delta\textbf{P}=\frac{1}{6}g_{(\alpha),AB}\biggl\{
\textbf{L}_{0}(\nabla\cdot\delta\textbf{L})+\delta\textbf{L}(\nabla\cdot\textbf{L}_{0})$$
$$-(\textbf{L}_{0}\cdot\nabla)\delta\textbf{L}-(\delta\textbf{L}\cdot\nabla)\textbf{L}_{0}$$
$$-\textbf{M}_{0}(\nabla\cdot\delta\textbf{M})-\delta\textbf{M}(\nabla\cdot\textbf{M}_{0})$$
\begin{equation}\label{MEEinAFMnc  P NONsimm lin}
+(\textbf{M}_{0}\cdot\nabla)\delta\textbf{M} +(\delta\textbf{M}\cdot\nabla)\textbf{M}_{0} \biggr\}
, \end{equation}
which includes the space dependence of the equilibrium vectors $\textbf{L}_{0}$ and $\textbf{M}_{0}$ on coordinates.

\begin{widetext}
\begin{equation}\label{MEEinAFMnc P perturb projections}
\left(
  \begin{array}{c}
    \delta P_{x}=0 \\
    \delta P_{y}=\frac{1}{6}g_{(\alpha),AB}\frac{1}{\imath\omega} \{ -2 (g_{0u}+\kappa +\imath\omega a)qL_{b}M_{b}\delta L_{z} +[0.5\kappa+g_{0u}+0.5\imath\omega a]L_{b}^{2}\partial_{x}\delta M_{z} +g_{(\alpha),AB}q^{2}L_{b}M_{c}L_{0y}\delta E_{z}\}\\
    \delta P_{z}=\frac{1}{6}g_{(\alpha),AB} \{\delta L_{z} \partial_{x}L_{0x} -L_{0x} \partial_{x}\delta L_{z} -\delta M_{z} \partial_{x}M_{0x} +M_{0x} \partial_{x}\delta M_{z}\} \\
  \end{array}
\right)
, \end{equation}
\end{widetext}
where $\delta P_{y}$ appears as a combination of $\delta L_{x}$ and $\delta L_{y}$, \emph{and} $\delta M_{x}$ and $\delta M_{y}$,
which are represented mainly via $\delta L_{z}$, $\delta M_{z}$, and $\delta E_{z}$.

The last term in $\delta P_{y}$ is proportional to $\delta E_{z}$ and appears in the nonresonance form.
Hence we are focused on the first two terms proportional to $\delta L_{z}$ and $\delta M_{z}$.
Further analysis of $\delta\textbf{P}$ requires expressions of $\delta L_{z}$ and $\delta M_{z}$ in terms of $\delta\textbf{E}$.

The spin density also responses on the perturbations of the magnetic field,
which can be represented in terms of the electric field perturbations
(see the discussion of the form of the dielectric response in Ref. \cite{Andreev 2025 05}
eqs. (15)-(21)).
Here, we point out features coming out of the megnetoelectric coupling.
Some of them can be hidden during observation due to standard magnetic effects, like the AFM resonance.
The alternating magnetic field leads to this effect as well as the magnetic part of the electromagnetic wave.
To some extend, we can expect to avoid this "noise" if the alternating electric field is used.

\subsection{Spin density perturbations}

Next, we present the consequences of the Landau--Lifshitz--Gilbert equations
considered in the nearest neighbours interaction approximation in the limit of the continuous medium \cite{Andreev 2026 03}
(see also \cite{Andreev 2025 Vestn} for more extended description of coefficients in the Landau--Lifshitz--Gilbert equations).
Methods of manipulation of projections of the small amplitude perturbations over the cycloidal equilibrium allowing to exclude the variable factors are described in Refs. \cite{Andreev 2025 09} and \cite{Andreev 2025 10}.

Here we demonstrate major contributions in evolution of $\delta L_{z}$
$$-\omega^{2}\delta L_{z}
=(g_{0u} +0.5 \imath\omega a)(0.5\kappa+0.5Aq^2+0.5\imath\omega a)L_{b}^2 \delta L_{z}$$
$$-g_{0u}AqL_{b}M_{b} \partial_{x}\delta M_{z}
+\imath\omega\frac{2}{3}g_{(\alpha),AB} qL_{b}M_{b} \delta E_{z}$$
\begin{equation}\label{MEEinAFMnc L lin appr evol}
- \frac{1}{3}g_{0u}g_{(\alpha),AB} qL_{b}L_{c}L_{0y}\delta E_{z}
, \end{equation}
and $\delta M_{z}$
$$-\omega^{2}\delta M_{z}
=(g_{0u}+0.5\kappa)L_{b}^{2} [0.5A\triangle\delta M_{z}+0.5\imath\omega a \delta M_{z}]$$
$$-A(g_{0u}+\kappa)qL_{b}M_{b}\partial_{x}\delta L_{z}
-\imath\omega\frac{1}{6}g_{(\alpha),AB} L_{b}^{2}\partial_{x}\delta E_{y}$$
\begin{equation}\label{MEEinAFMnc M lin appr evol}
-\frac{1}{2}A q^2 g_{(\alpha),AB} q L_{b} L_{c} M_{0y} \delta E_{z}
, \end{equation}
which appears at the neglecting of high derivatives of $\delta L_{z}$, $\delta M_{z}$, and $\delta \textbf{E}$ on space coordinate.
We also neglect $M_{b}$ in comparison with $L_{b}$.
The contribution of interactions responsible for the equilibrium weak ferromagnetism is also dropped in the dynamical equations.

Equations (\ref{MEEinAFMnc L lin appr evol}) and (\ref{MEEinAFMnc M lin appr evol}) show
that for the magnetoelectric effect of chosen type
there is dielectric response is nonzero for two projections of the electric field $\delta E_{y}$ and $\delta E_{z}$.
Moreover, the evolution of $\delta L_{z}$ responsible for the AFM resonance responses on $\delta E_{z}$ only.
The second term on rhs of equation (\ref{MEEinAFMnc L lin appr evol}) shows the contribution of $\delta M_{z}$ in $\delta L_{z}$ dynamics.
So, we have a mechanism of influence of $\delta E_{y}$ on the AFM resonance due to the coupling of two spin waves in AFM.
Similarly, we have influence of $\delta L_{z}$ on $\delta M_{z}$ dynamics
(see the second term on the rhs of equation (\ref{MEEinAFMnc M lin appr evol}))
completing the selfconsistent dynamics of the spin waves.

Described coupling between two spin waves happens due to simultaneous influence of two factors:
weak ferromagnetism (nonzero value of $M_{b}$) and the cycloidal structure $q\neq0$.

It is essential to point out the last term in equation (\ref{MEEinAFMnc L lin appr evol})
and the last term in equation (\ref{MEEinAFMnc M lin appr evol}) contains periodic coefficient
$L_{0y}=L_{c}\sin(qx)$ and $M_{0y}=-M_{c}\cos(qx)$, correspondingly.
If we drop the second term on rhs of equation (\ref{MEEinAFMnc L lin appr evol})
we obtain algebraic equation for $\delta L_{z}$ and express $\delta L_{z}$ via $\delta E_{z}$
for the further substitution in the polarization perturbations (\ref{MEEinAFMnc P perturb projections}).
Equation for $\delta M_{z}$ (\ref{MEEinAFMnc M lin appr evol}) shows more complex behavior.
The major term on the rhs contains the second derivative of $\delta M_{z}$:
$0.5A\triangle\delta M_{z}$.
We can use the slowly varying amplitude approach for the introduction of the waves vector
since we consider the long-wavelength approximation and dropped higher derivatives in equations
(\ref{MEEinAFMnc L lin appr evol}) and (\ref{MEEinAFMnc M lin appr evol}):
$0.5A\triangle\delta M_{z}\approx -0.5Ak^2\delta M_{z}$.
Hence we can represent
$-(g_{0u}+0.5\kappa)L_{b}^{2} [0.5A\triangle\delta M_{z}+0.5\imath\omega a \delta M_{z}]$
as
\begin{equation}\label{MEEinAFMnc omega l def}
0.5(g_{0u}+0.5\kappa)L_{b}^{2} [Ak^2-\imath\omega a]\delta M_{z}\equiv\omega_{eff,low}^{2}\delta M_{z}\equiv\omega_{l}^{2}\delta M_{z}.
\end{equation}
It leads to the approximate expression of $\delta M_{z}$ in terms of perturbations of the electric field
\begin{equation}\label{MEEinAFMnc delta M z NO h}
\delta M_{z}=\frac{1}{2}g_{(\alpha),AB}\frac{\imath\omega\frac{1}{3} L_{b}^{2}\partial_{x}\delta E_{y}+A q^2 \cdot q L_{b} L_{c} M_{0y} \delta E_{z}}{\omega^{2}-\omega_{l}^{2}},
\end{equation}

We also introduce the characteristic frequency for $\delta L_{z}$
\begin{equation}\label{MEEinAFMnc omega h def}
\omega_{h}^{2}\equiv\omega_{eff,high}^{2}\equiv-0.5(g_{0u} +0.5 \imath\omega a)(\kappa+Aq^2+\imath\omega a)L_{b}^2
. \end{equation}
Hence, we able to present the perturbations of the AFM vector in terms of perturbations of the electric field
(in this case we do not need to use the varying amplitude approach)
\begin{equation}\label{MEEinAFMnc delta L z NO h}
\delta L_{z}=\frac{1}{3}g_{(\alpha),AB}\frac{-\imath\omega 2 qL_{b}M_{b} \delta E_{z}+g_{0u} qL_{b}L_{c}L_{0y}\delta E_{z}}{\omega^{2}-\omega_{h}^{2}}.
\end{equation}

Let us to point out that
the contribution of the out of plane (relatively to cycloid) electric field $\delta E_{z}$ appears due to the cycloid structure
since corresponding terms in equations (\ref{MEEinAFMnc delta M z NO h}) and (\ref{MEEinAFMnc delta L z NO h})
are proportional to the cycloid wave vector $q$.

In order to get the stable low (high) frequency wave corresponding to
$\omega_{l}$ (\ref{MEEinAFMnc omega l def})
[$\omega_{h}$ (\ref{MEEinAFMnc omega h def}) ]
we need positive left-hand side of equation (\ref{MEEinAFMnc omega l def})
[of equation (\ref{MEEinAFMnc omega h def})].
It gives $g_{0u}>0$, $A>0$, but $\kappa<0$ as it is known for the easy-plane samples
(see Sec. 4.1.2 in Ref. \cite{Andreev 2025 10}).
Including this information we can rewrite approximate expressions
as follows
$\omega_{l}^{2}=0.5(g_{0u}-0.5\mid\kappa\mid)Ak^2 L_{b}^{2}$,
and
$\omega_{h}^{2}=g_{0u} (0.5\mid\kappa\mid-0.5Aq^2)L_{b}^2$.

\emph{No hybridization of waves regime}

We neglected the hybridization of two spin waves.
Next, we focus on the electric field perturbations located in plane of cycloid
$\delta E_{z}=0$, but $\delta E_{y}\neq 0$.
It leads to $\delta L_{z}=0$ and gives no AFM resonance in the dielectric response.

Described approximations leads to the simplified form of polarization presented in terms of perturbations of the magnetization vector only
\begin{equation}\label{MEEinAFMnc P perturb projections Simplified}
\left(
  \begin{array}{c}
    \delta P_{x}=0 \\
    \delta P_{y}=\frac{1}{6}g_{(\alpha),AB}\frac{1}{\imath\omega}  [0.5\kappa+g_{0u}+0.5\imath\omega a]L_{b}^{2}\partial_{x}\delta M_{z} \\
    \delta P_{z}=-\frac{1}{6}g_{(\alpha),AB} (\partial_{x}M_{0x}) \delta M_{z} \\
  \end{array}
\right)
. \end{equation}
Made approximations allows to give a simple form of the perturbations of the magnetization vector as well
\begin{equation}\label{MEEinAFMnc }
\delta M_{z}=\frac{1}{6}g_{(\alpha),AB}\imath\omega\frac{ L_{b}^{2}\partial_{x}\delta E_{y}}{\omega^{2}-\omega_{l}^{2}}.
\end{equation}
All of these lead to the polarization perturbations presented in terms of the electric field perturbations
\begin{equation}\label{MEEinAFMnc P perturb projections Simplified Substituted and Dropped}
\left(
  \begin{array}{c}
    \delta P_{x}=0 \\
    \delta P_{y}=\frac{1}{6^2}g_{(\alpha),AB}^2 \frac{ [g_{0u}+0.5\kappa+0.5\imath\omega a]}{\omega^{2}-\omega_{l}^{2}}  L_{b}^{4}
    \partial_{x}^2\delta E_{y} \\
    \delta P_{z}=\frac{1}{6^2}g_{(\alpha),AB}^2 \frac{ (-\imath\omega)}{\omega^{2}-\omega_{l}^{2}}L_{b}^{2}(\partial_{x}M_{0x})\partial_{x}\delta E_{y}  \\
  \end{array}
\right)
, \end{equation}
which represents the dynamical electric susceptibility tensor $\kappa^{\alpha\beta}$ in the chosen regime.
Here we also neglected $M_{0x} \partial_{x}\delta M_{z}\approx \imath k M_{0x} \delta M_{z}$ in compare with
$\delta M_{z} \partial_{x}M_{0x}\approx \delta M_{z} qM_{0y}$.
Compare $\delta P_{y}$ and $\delta P_{z}$ in equation
(\ref{MEEinAFMnc P perturb projections Simplified Substituted and Dropped}).
Expression for $\delta P_{y}$ has no signatures of the weak ferromagnetism,
and this dielectric response is possible for more simple structures with $\textbf{M}_{0}=0$.
Presence of the second derivative of the electric field $\partial_{x}^2\delta E_{y}\sim k^2\delta E_{y}$ notes weakness of the response
since we deal with the long-wavelength limit $k\ll q$.
On the other hand, $\delta P_{z}$ shows explicit dependence both on the weak ferromagnetism in the equilibrium state $M_{b}\neq 0$
and the cycloidal structure $q\neq 0$
via factor $\partial_{x}M_{0x}=qM_{b}\cos(qx)$.

Further approximation and comparison of the response near the resonance shows
\begin{equation}\label{MEEinAFMnc P perturb projections Simplified Substituted and Dropped NEAR resonance}
\left(
  \begin{array}{c}
    \delta P_{x}=0 \\
    \delta P_{y}=r \biggl[0.5 k^2 a^2 -\frac{\imath a k[g_{0u}+0.5\kappa]}{\sqrt{0.5(g_{0u}-0.5\mid\kappa\mid)A}L_{b}}\biggr]
    L_{b}^{2} \delta E_{y} \\
    \delta P_{z}=-r \cdot\imath a qk M_{b} \delta E_{y}  \\
  \end{array}
\right)
, \end{equation}
where parameter $r=(g_{(\alpha),AB}^2/6^2)(0.5a^2(g_{0u}+0.5\kappa))^{(-1)}$
is used to combine the repeating combination.
The following approximate replacements are made
$\partial_{x}\delta E_{y} \sim \imath k \delta E_{y}$
and
$\partial_{x}^2\delta E_{y} \sim - k^2 \delta E_{y}$.
The maximal value of oscillating $\partial_{x}M_{0x}$ is included as $-qM_{b}$ with the proper sign.
It leads to
$$\frac{Im\delta P_{y}}{Im\delta P_{z}}=\frac{ \sqrt{[g_{0u}+0.5\kappa]L_{b}}}{\sqrt{0.5A q ^2 L_{b}}}\frac{L_{b}}{M_{b}}\gg 1,$$
while $0.5A q ^2 L_{b}$ is a relatively large frequency.
We use $A\approx 2\cdot10^5$ cm$^{3}$/g, and $n_{0}=10^{22}$ cm$^{-3}$.
Here we also refer to \cite{Dong AinP 15} Fig. 3 to use $q\approx 10^6$ cm$^{-1}$.
It gives us the characteristic low frequency
$\omega_{ch}=Aq^2L_b\approx 10^{12}$ s$^{-1}$.

An estimation of $g_{(\alpha),AB}$ via the constant of the exchange interaction $A$ based on the spin-current model
can be found in Ref. \cite{Andreev 2025 05} (see eq. 8).
We also use it here $g_{(\alpha),AB}/3=-g_{(\alpha)}/3=0.5 A(\gamma/c)=0.4\cdot10^2$ CGS.
To extract the electric susceptibility from equation
(\ref{MEEinAFMnc P perturb projections Simplified Substituted and Dropped})
we also use the slowly varying amplitude approach and write
$\partial_{x}^2\delta E_{y}\sim k^2\delta E_{y}$ and $\partial_{x}\delta E_{y}\sim \imath k\delta E_{y}$,
correspondingly.

\emph{Hybridization of waves regime}

Equations (\ref{MEEinAFMnc L lin appr evol}) and (\ref{MEEinAFMnc M lin appr evol}) show relatively strong hybridization of spin waves,
which is formally neglected above.
It appears for the equilibrium uniting the cycloidal structure of the AFM vector (\ref{MEEinAFMnc L eq cycloid})
and nonzero (also cycloidal) magnetization vector (\ref{MEEinAFMnc M eq cycloid}).
Assumption about strong coupling follows from the fact that corresponding terms are proportional to
products of constants describing the exchange interaction $g_{0u}A$ and $(g_{0u}+\kappa)A$.

This coupling leads to the appearance of the AFM resonance in the dynamical electric susceptibility
in response to the electric field acting in plane of the cycloid $\delta E_{y}$,
but being perpendicular to the direction of the spin wave propagation and the equilibrium cycloid "propagation".
Basically we compare the propagation of two electromagnetic waves,
propagating in the direction of cycloid (so their wave vectors are parallel $\textbf{k}\parallel \textbf{q}$).
So, we have two linear polarizations for the electromagnetic wave
(in plane of the cycloid $\delta E_{y}$ and out of plane $\delta E_{z}$).

If we consider the collinear AFM we find no such coupling.
However, the weak coupling between two spin waves still exist,
but it is related to
the Dzylaoshinskii-Moriya interaction in the form related to two partial ligand shifts
and OASEI (see Ref. \cite{Andreev 2025 12}).

We are consciously consider the properties of the electric susceptibility,
since the refractive index can contain some extra information from the magnetization dynamics
(see for instance Ref. \cite{Andreev 2025 05}).
Is it possible to deal with the alternating electric field,
which is not a part of the electromagnetic wave to avoid the action of the magnetic field?
Since the magnetic field causes the AFM resonance covering features described above.

This coupling is mostly interesting in appearance of $\delta L_z$
and it further contribution in the electric polarization,
which also shows the coupling caused shifts of resonances:
\begin{equation}\label{MEEinAFMnc }
\delta L_z=
\frac{-\frac{1}{6}g_{(\alpha),AB}\omega k g_{0u}A q L_b^3 M_b \partial_{x}\delta E_{y}}{(\omega^{2}-\omega_{l}^{2})(\omega^{2}-\omega_{h}^{2}) +g_{0u}(g_{0u}+\kappa)A^2 q^2 k^2 L_b^2 M_b^2}, \end{equation}

We can simplify it for the area of the low-frequency resonance
\begin{equation}\label{MEEinAFMnc delta L_z appr with further appr}
\delta L_z=
\frac{\frac{1}{6}g_{(\alpha),AB}\omega k A q L_b M_b \partial_{x}\delta E_{y}}{ 0.5 \mid\kappa\mid(\omega^{2}-\omega_{l}^{2}) -(g_{0u}-\mid\kappa\mid)A^2 q^2 k^2 M_b^2}  , \end{equation}
with $\mid\kappa\mid-A q^2\approx \mid\kappa\mid$ as an additional approximation.
Here we get $(g_{0u}-\mid\kappa\mid)<g_{0u}$, but we can assume that $g_{0u}$ and $\mid\kappa\mid$ are comparable.
Hence, we can compare $\omega^{2}-\omega_{l}^{2}$ and $A^2 q^2 k^2 M_b^2$
in the denominator,
where $\omega^{2}-\omega_{l}^{2}$
$=-\imath\omega a 0.5(g_{0u}+0.5\kappa)L_{b}^{2}$
$=-\imath\sqrt{0.5(g_{0u}-0.5\mid\kappa\mid)A} a 0.5(g_{0u}+0.5\kappa) k L_{b}^{3}$
is imaginary, but its module larger for the long-wavelength limit,
and also due to $M_b^2/L_b^2\ll 1$.

It leads to the additional contribution in the dynamical polarization
\begin{equation}\label{MEEinAFMnc P perturb projections L}
\left(
  \begin{array}{c}
    \delta P_{x,L}=0 \\
    \delta P_{y,L}=-\frac{1}{3}g_{(\alpha),AB}\frac{1}{\imath\omega}  (g_{0u}+\kappa +\imath\omega a)qL_{b}M_{b}\delta L_{z} \\
    \delta P_{z,L}=\frac{1}{6}g_{(\alpha),AB} (\partial_{x}L_{0x})\delta L_{z}   \\
  \end{array}
\right)
, \end{equation}
where
we assumed $L_{0x} \partial_{x}\delta L_{z}\ll \delta L_{z} \partial_{x}L_{0x}$
and corresponding $\delta L_{z}$ is in equation (\ref{MEEinAFMnc delta L_z appr with further appr}).

If we consider two contributions in $\delta P_{y}$
we find $\delta P_{y,M}\sim L_{b}^4$, while $\delta P_{y,L}\sim M_{b}^2 L_{b}^2$,
hence the contribution of $\delta M_z$ is major in the low frequency limit.
Next, consider $\delta P_{z}$,
we find $\delta P_{z,M}\sim M_b L_{b}^2$ and $\delta P_{z,L}\sim M_b L_{b}^2$,
but there is a spatial shift in the coefficients,
so they do not cancel each other.

As an additional comparison we mention Ref. \cite{Aupiais npj QM 18},
where the dielectric response of the magnetic material is considered for a realistic and complex structure.
Ref. \cite{Aupiais npj QM 18} is mostly focused on the response related to the electromagnons.
While here we focus on the magnetic dynamics and its consequence only,
we also deal with the low frequency area,
where the electromagnon resonance is located.

\section{Conclusion}

Purely magnetic dynamics leading to the dielectric response has been considered for the AFM multiferroics
with the cycloidal structure including the weak ferromagnetism.
The magnetoelectric effect of spin origin has been considered for the noncollinear parts of spins.

While the cycloid has been placed in XY plane and "propagates" parallel x-axis it responses on the y- and z-projections of the dynamical electric field.
If we neglect the coupling/hybridization between two spin waves
(which is relatively strong due to the weak ferromagnetism in the cycloidal structure)
we obtain that
high-frequency wave response (corresponding to the AFM resonance) happens for the z-projections of the dynamical electric field only.
We are interested in the low-frequency response,
hence we focus our attention on the y-projections of the dynamical electric field $\delta E_{y}$.
It produces two projections of the dynamical polarization $\delta P_{y}$ and $\delta P_{z}$.
Our estimation shows that
the y-projection of the dynamical polarization $\delta P_{y}$ gives major effect.

If we include the hybridization between spin waves it gives an extra contributions in $\delta P_{y}$ and $\delta P_{z}$.
Regarding the low-frequency resonance we have found that additional contribution in $\delta P_{y}$ is rather small,
while addition effect in $\delta P_{z}$ is comparable with the main effect considered in the no hybridization limit.

Presence of $\delta P_{z}$,
due to considered mechanism of the polarization formation,
appears to be a dynamical indicator of the weak ferromagnetism in the AFM cycloid.
It also appears as the result of tensor dielectric response on the dynamical electric field $\delta E_{y}$.
This low frequency dynamical response is essential at the analysis of the electromagnon resonances.


\emph{DATA AVAILABILITY}:

Data sharing is not applicable to this article as no new data were
created or analyzed in this study, which is a purely theoretical one.


\emph{Acknowledgements}:
The work is supported by the Russian Science Foundation under the grant No. 25-22-00064.



\end{document}